\documentclass[onecolumn,pra,aps,amssymb]{revtex4}
\usepackage{hyperref}
\usepackage{graphicx}
\usepackage{amsmath,amssymb,amsfonts}
\usepackage{algorithm}
\usepackage{multirow}
\usepackage{verbatim}
\usepackage{url}
\usepackage{comment}
\usepackage{mathtools}
\usepackage{epsf,pgf,graphicx}
\usepackage{color}
\usepackage{tikz,pgf}

\begin{document}
\newcommand{\ket}[1]{\ensuremath{\left|#1\right\rangle}}
\newcommand{\bra}[1]{\ensuremath{\left\langle#1\right|}}
\newcommand\floor[1]{\lfloor#1\rfloor}
\newcommand\ceil[1]{\lceil#1\rceil}

%%%%%%%%%%%%%%%%%%%%% Publisher's Area please ignore %%%%%%%%%%%%%%
%\catchline{}{}{}
%%%%%%%%%%%%%%%%%%%%%%%%%%%%%%%%%%%%%%%%%%%%%%%%%%%%%%%%%%%%%%%%%%%

\title{Measurement Device Independent Quantum Private Query with Qutrits}

\author{Sarbani Roy$^1$\footnote{sarbani16roy@gmail.com}
, Arpita Maitra$^2$\footnote{arpita76b@gmail.com}
, Sourav Mukhopadhyay$^3$\footnote{msourav@gmail.com}}
\address{$^{1,3}$ Department of Mathematics,\\
$^2$ Centre for Theoretical Studies, \\
Indian Institute of Technology Kharagpur--721302, India}

%\address{Department of Mathematics, Indian Institute of Technology Kharagpur--721302, India}

%\begin{history}
%\received{Day Month Year}
%\revised{Day Month Year}
%\accepted{Day Month Year}
%\comby{(xxxxxxxxxx)}
%\end{history}

\begin{abstract}
Measurement Device Independent Quantum Private Query (MDI QPQ) with qutrits is presented.
We compare the database security and client's privacy in MDI QPQ for qubits with qutrits. For some instances, we observe that qutrit will provide better security for database than qubit. However, when it comes to the question of client's privacy we have to take additional measures in case of qutrit. Hence we conclude that though in case of Quantum Key Distribution (QKD) higher dimension provides better security  but in case of QPQ this is not obvious.
\end{abstract}
\maketitle
%\keywords{Quantum Private Query; Qutrit; Measurement Device Independence; Security Analysis
%}
\section{Introduction}
\label{intro}
The very first protocol of quantum cryptography was Bennett and Brassard's BB84~\cite{bb84} key distribution protocol. In these thirty years, the field is progressing gradually. Though the main focus of quantum cryptography is quantum key distribution (QKD),~\cite{bb84,b92,SARG,jo} which is to establish a secret key between two distant parties, several other quantum cryptographic primitives have also been proposed and actively studied, such as quantum secret sharing (QSS),~\cite{hbb,wmy} quantum secure direct communication, (QSDC)~\cite{ll,DLL,zhang} quantum private query (QPQ),~\cite{GLM,jakobi,Gao,Yang,zhao} quantum position verification (QPV)~\cite{glw} and so on.

Measurement-device-independent (MDI) technique is a very powerful and important technique in quantum cryptography. It allows the secure quantum communication using imperfect measurement devices. This has been done for QKD, QPQ with qubit. Recently, MDI-QSDC~\cite{zhou} has been proposed.

Let us consider the situation where Bob is a database provider and Alice is a user. Now Alice wants to know a certain element from that database without providing any information about her query to Bob. On the other hand, Bob tries to resist Alice to know any other elements from the database except her query. The first one is called user or client's privacy whereas the second one is known as database security. This problem is a variant of Symmetric Private Information Retrieval (SPIR). The hardness assumptions which are exploited in SPIR protocols~\cite{ps,ko} are proven to be vulnerable in quantum domain. This is why the researcher searched for some SPIR protocols which can resist the quantum adversary, an adversary having unbounded power of computation.
In~\cite{lo1} it has been shown that a prefect quantum SPIR protocol is impossible. However, Giovannetti et al.~\cite{GLM} came out with a variant of this and called it Quantum Private Query (QPQ).  Some relaxations in the security notions had made it feasible in quantum paradigm.

However, the protocol suggested by Giovannetti et al. was purely theoretical and difficult for implementation. Jacobi et al.~\cite{jakobi} for the first time proposed a practical QPQ protocol based on SARG04~\cite{SARG} QKD protocol. Using SARG04 QKD protocol, an asymmetric key which is used to encrypt the whole database, is distributed between Bob and Alice.  According to the protocol, $\frac{1}{4}$ portion of the key is known to Alice. Gao et al.~\cite{Gao} proposed a flexible QPQ protocol (formally known as GLWC protocol) by generalizing Jacobi's protocol. The difference is that in GLWC protocol Alice knows $\frac{\sin^2\theta}{2}$ portion of the key, where the parameter $\theta\in(0,\frac{\pi}{2})$ is chosen by Bob based on the required amount of security. For $\theta<\frac{\pi}{4}$, GLWC protocol provides better database security but there is a high probability to guess the address of Alice's query. To solve this issue, Yang et al.~\cite{Yang} proposed an entanglement based QPQ protocol based on B92~\cite{b92} QKD Protocol.

Very recently, Zhao et al.~\cite{zhao} has developed a
detector-blinding attack by dishonest Bob on Jacobi and GLWC
protocol and shown that how detector's side channel attack breaks
the user privacy completely. To remove all such type of side channel
attacks, they then proposed a Measurement Device Independent (MDI)
QPQ protocol exploiting MDI QKD~\cite{lo2,pirandola,Maitra17}.

 In MDI QKD, Alice and Bob send their encoded photons to an untrusted third party (UTP) Charlie who performs the Bell state measurement (BSM) and announces the measurement result to Alice and Bob. Depending on Charlie's announcement,  Alice and Bob can find a correlation between their photons and thus can establish a secret key between themselves. In MDI QPQ~\cite{zhao}, this idea has been exploited to establish an asymmetric key between Bob and Alice. Database privacy remains same as previous QPQ protocols~\cite{Gao,Yang}. They proved the loss tolerance user privacy under some specific attack models.

Recently, Jo and Son~\cite{jo} have proposed a MDI QKD protocol using qutrits. It is expected that a higher dimensional state can carry more information per quanta than a qubit. And hence they have noticed an enhancement in key rate compared to qubit MDI QKD.

 Motivated by this, we try to understand if it is indeed the case for all cryptographic primitives. In this direction, we explore MDI QPQ for qutrit. We observe that for some instances if we go for higher dimension, then it is possible to obtain better database security. However, in such cases to protect the user privacy we have to consider some additional measures. That indicates that in case of qutrit we can not optimize both database security and user privacy simultaneously. The security issues are very much protocol specific. It disproved our conjecture that higher dimension always provides better security.
\section{Measurement Device Independent Quantum Private Query using Qutrits }
In this section we first accumulate some assumptions necessary for this protocol. Then we will enumerate MDI-QPQ for qutrit.% In~\cite{zhao} these assumptions are implicit.

\subsection{Necessary Assumptions}
\begin{enumerate}
\item In an honest run of the protocol, Alice and Bob do not deviate from the actions they are supposed to do.
\item In an honest run of the protocol, Alice or Bob does not collude with Charlie, the untrusted third party.
\end{enumerate}

\subsection{MDI-QPQ with Qutrit} In case of qutrit the dimension of Hilbert space is three. The general form of a qutrit $\ket{\psi}$ can be expressed as
\begin{equation}
\ket{\psi} = \cos \gamma_1 \ket{0} + \sin \gamma_1 \cos \gamma_2 \ket{1} + \sin \gamma_1 \sin \gamma_2 \ket{2},
\end{equation}
where $\gamma_1, \gamma_2$ are parameters such that $ 0 \leq \gamma_1, \gamma_2 \leq \frac{\pi}{2} $.
\\
By applying a specific unitary transformation on the computational
basis vectors $\{\ket{0}, \ket{1}, \ket{2}\}$, one can get a set of orthonormal basis vectors which includes the state of $(1)$. Following are the examples of such basis states.
\begin{eqnarray*}
    \ket{0'} & =& U(\ket{0})\\
    & =& \cos \gamma_1 \ket{0} + \sin \gamma_1 \cos \gamma_2 \ket{1} + \sin \gamma_1 \sin \gamma_2 \ket{2}\\
    \ket{1'} & = &U(\ket{1}) \\
    &=& -\sin \gamma_1 \ket{0} + \cos \gamma_1 \cos \gamma_2 \ket{1} + \cos \gamma_1 \sin \gamma_2 \ket{2}\\
    \ket{2'} &=& U(\ket{2})\\
    & = &-\sin \gamma_2 \ket{1} + \cos \gamma_2 \ket{2}\\
    \end{eqnarray*}

There are nine maximally entangled states of the three dimensional bipartite system. We define $\{ \ket{\phi_i} \}$ as the set of three-dimensional maximally entangled states, where $i \in \{0,1,\ldots,8\}$, and each state is described as

\begin{equation}
\ket{\phi_{3k+l}} = \frac{1}{\sqrt 3}\sum_{m=0}^{2} \omega^{ml}\ket{m+k,m}
\end{equation}
where $ k,l \in \{ 0,1,2\}$ and $\omega = e^{\frac{2\pi i}{3}}$. We
omit (mod 3) from all indices for simplification. Then the
three-dimensional Bell state measurement ($3d$-BSM) is defined as a
set of projections $\{ \hat{P_i} =
\ket{\phi_i}\bra{\phi_i} \}$.

In MDI QPQ, Charlie performs the Bell state measurement (BSM) with the photons coming from Bob and Alice in the motivation towards establishing an asymmetric key between Alice and Bob. Charlie may play the role of an eavesdropper.

However, the QPQ protocol is viewed as a mistrustful cryptographic primitive. Hence, we need not to consider an outsider (Charlie) as an eavesdropper. Either Bob or Alice may behave as an adversary. We will analyze the security issues in this initiative.

\ \\
{\bf{Protocol}}:
\begin{enumerate}

\item Bob sends, uniformly at random, one of the six polarized states $\ket{0}$, $\ket{1}$, $\ket{2}$, $\ket{0'}$, $\ket{1'}$, $\ket{2'}$ to Charlie, where $\gamma_1, \gamma_2 \in (0, \frac{\pi}{2})$. The rectilinear basis $\{\ket{0}, \ket{1}, \ket{2}\}$ encodes to key bit $0$ and the basis$\{\ket{0'}, \ket{1'}, \ket{2'}\}$ encodes to the key bit $1$.
\item Alice sends, uniformly at random, one of the six polarized states $\ket{0}$, $\ket{1}$, $\ket{2}$, $\ket{0'}$, $\ket{1'}$, $\ket{2'}$ to Charlie. The parameters $\gamma_1, \gamma_2$ are same as Bob. They should have prior discussion about this.
\item Charlie performs the BSM and declares the results publicly. Alice records the measurement result. In fact, she needs to identify only the Bell state $\ket{\phi_0}$. Theoretical probabilities for obtaining $\ket{\phi_0}$ for different combinations are shown in the Table $1$.
\item For each trial Charlie obtained a Bell state $\ket{\phi_0}$, Bob announces a trit 0 to Alice if he has sent $\ket{0}$ or $\ket{0'}$, 1 if he has sent $\ket{1}$ or $\ket{1'}$ and 2 if he has sent$\ket{2}$ or $\ket{2'}$.
\item Depending on the state Alice has sent to Charlie and the declaration
 of Bob she will try to guess the key bit.
 \item After the key establishment Alice and Bob go for error correction to check the noise in the channel.
\item If Alice knows the $j$th bit of the key $K$ and wants to know the $i$th element of the database, she declares the integer $s=j-i$.
Bob shifts $K$ by $s$ and hence gets a new key, say $K_0$.
\item Bob encrypts his database by this new key $K_0$ with one-time pad and sends the encrypted database to Alice.
 Alice decrypts the value with her $j$th key bit and gets the required element of the database.
\end{enumerate}

We now discuss how Alice obtains a conclusive raw key bit.

\begin{enumerate}

\item If Bob has sent the state $\ket{0}$ and announced the trit 0, according to Table $1$, Alice can identify Bob's state $\ket{0}$ and thus the raw key bit 0 with certainty only if she has prepared the state $\ket{1'}$.
\item If Bob has sent the state $\ket{0'}$ and announced the trit 0, according to Table $1$, Alice can identify Bob's state $\ket{0'}$ and thus the raw key bit 1 with certainty only if she has prepared the state $\ket{1}$ or $\ket{2}$.
\item If Bob has sent the state $\ket{1}$ and announced the trit 1, according to Table $1$, Alice can identify Bob's state $\ket{1}$ and thus the raw key bit 0 with certainty only if she has prepared the state $\ket{0'}$ or $\ket{2'}$.
\item If Bob has sent the state $\ket{1'}$ and announced the trit 1, according to Table $1$, Alice can identify Bob's state $\ket{1'}$ and thus the raw key bit 1 with certainty only if she has prepared the state $\ket{0}$ or $\ket{2}$.
\item If Bob has sent the state $\ket{2}$ and announced the trit 2, according to Table $1$, Alice can identify Bob's state $\ket{2}$ and thus the raw key bit 0 with certainty only if she has prepared the state $\ket{0'}$ or $\ket{1'}$.
\item If Bob has sent the state $\ket{2'}$ and announced the trit 2, according to Table $1$, Alice can identify Bob's state $\ket{2'}$ and thus the raw key bit 1 with certainty only if she has prepared the state $\ket{1}$.

\end{enumerate}

\begin{table*}[htbp]
\centering
{\scriptsize
\begin{tabular}{|c|c|c|c|c|c|c|c|}
\hline
 & \multicolumn{6}{c}{Bob} & \\
\cline{2-8}
& & $\ket{0}$ & $\ket{1}$ & $\ket{2}$ & $\ket{0'}$ & $\ket{1'}$ & $\ket{2'}$\\
\hline
\multirow{6}{*}{Alice} & $\ket{0}$ & $\frac{1}{3}$ & 0 & 0 & $\frac{1}{3}{\cos^2 \gamma_1}$ & $\frac{1}{3}{\sin^2 \gamma_1}$ & 0\\
\cline{2-8}
            & $\ket{1}$ & 0 & $\frac{1}{3}$ & 0 & $\frac{1}{3}{\sin^2 \gamma _1 \cos^2 \gamma_2}$ & $\frac{1}{3}{\cos^2 \gamma_1 \cos^2 \gamma_2}$ & $\frac{1}{3}{\sin^2 \gamma_2}$\\
\cline{2-8}
            & $\ket{2}$ & 0 & 0 & $\frac{1}{3}$ & $\frac{1}{3}{\sin^2 \gamma _1 \sin^2 \gamma_2}$ & $\frac{1}{3}{\cos^2 \gamma_1 \sin^2 \gamma_2}$ & $\frac{1}{3}{\cos^2 \gamma_2}$\\
\cline{2-8}
            & $\ket{0'}$ & $\frac{1}{3}{\cos^2 \gamma_1}$ & $\frac{1}{3}{\sin^2 \gamma _1 \cos^2 \gamma_2}$ & $\frac{1}{3}{\sin^2 \gamma _1 \sin^2 \gamma_2}$ & $\frac{1}{3}$ & 0 & 0 \\
\cline{2-8}
            & $\ket{1'}$ & $\frac{1}{3}{\sin^2 \gamma_1}$ & $\frac{1}{3}{\cos^2 \gamma_1 \cos^2 \gamma_2}$ &  $\frac{1}{3}{\cos^2 \gamma_1 \sin^2 \gamma_2}$ & 0 & $\frac{1}{3}$ & 0 \\
\cline{2-8}
            & $\ket{2'}$ & 0 &  $\frac{1}{3}{\sin^2 \gamma_2}$ & $\frac{1}{3}{\cos^2 \gamma_2}$ & 0 & 0 & $\frac{1}{3}$\\
\hline
\end{tabular}}
\caption{Theoretical probabilities for obtaining Bell state $\ket{\phi_0}$ for different combination of states}
\end{table*}

Note that Table $1$ is not normalized.  To normalized it we have to
divide every elements of the table by a fraction $\frac{2}{3}$.
Table $2$ cumulates the normalized elements of Table $1$.
\begin{table*}[htbp]
\centering
{\scriptsize
\begin{tabular}{|c|c|c|c|c|c|c|c|}
\hline
 & \multicolumn{6}{c}{Bob} & \\
\cline{2-8}
& & $\ket{0}$ & $\ket{1}$ & $\ket{2}$ & $\ket{0'}$ & $\ket{1'}$ & $\ket{2'}$\\
\hline
\multirow{6}{*}{Alice} & $\ket{0}$ & $\frac{1}{2}$ & 0 & 0 & $\frac{1}{2}{\cos^2 \gamma_1}$ & $\frac{1}{2}{\sin^2 \gamma_1}$ & 0\\
\cline{2-8}
            & $\ket{1}$ & 0 & $\frac{1}{2}$ & 0 & $\frac{1}{2}{\sin^2 \gamma _1 \cos^2 \gamma_2}$ & $\frac{1}{2}{\cos^2 \gamma_1 \cos^2 \gamma_2}$ & $\frac{1}{2}{\sin^2 \gamma_2}$\\
\cline{2-8}
            & $\ket{2}$ & 0 & 0 & $\frac{1}{2}$ & $\frac{1}{2}{\sin^2 \gamma _1 \sin^2 \gamma_2}$ & $\frac{1}{2}{\cos^2 \gamma_1 \sin^2 \gamma_2}$ & $\frac{1}{2}{\cos^2 \gamma_2}$\\
\cline{2-8}
            & $\ket{0'}$ & $\frac{1}{2}{\cos^2 \gamma_1}$ & $\frac{1}{2}{\sin^2 \gamma _1 \cos^2 \gamma_2}$ & $\frac{1}{2}{\sin^2 \gamma _1 \sin^2 \gamma_2}$ & $\frac{1}{2}$ & 0 & 0 \\
\cline{2-8}
            & $\ket{1'}$ & $\frac{1}{2}{\sin^2 \gamma_1}$ & $\frac{1}{2}{\cos^2 \gamma_1 \cos^2 \gamma_2}$ &  $\frac{1}{2}{\cos^2 \gamma_1 \sin^2 \gamma_2}$ & 0 & $\frac{1}{2}$ & 0 \\
\cline{2-8}
            & $\ket{2'}$ & 0 &  $\frac{1}{2}{\sin^2 \gamma_2}$ & $\frac{1}{2}{\cos^2 \gamma_2}$ & 0 & 0 & $\frac{1}{2}$\\
\hline
\end{tabular}}
\caption{Theoretical probabilities for obtaining Bell state $\ket{\phi_0}$ for different combination of states after normalization}
\end{table*}
\section{Comparison of the security issues for qubit with qutrit}
In this section we compare the security issues of MDI QPQ protocol
for qubit with qutrit assuming one of  the parties is dishonest.
\subsection{Database Security}
In case of database security we assume that Alice is dishonest. Her
goal is to exact more element from the database except her query.
This will be possible if she can increase her success probability to
guess a key bit than what is suggested by the protocol. In this
subsection we will see how we can reduce the success probability of
Alice and hence can enhance the database security exploiting qutrit.

From Table $2$ we now calculate the success probability of Alice to
guess a raw key bit of Bob. Let $A$ be a random variable that Alice
has guessed about $B$ (random variable possessed by Bob). Thus the
probability of success can be written as

 {\scriptsize
\begin{eqnarray*}
 \Pr (A=B) & =& \Pr(A=0, B=0) + \Pr(A=1, B=1)\\
                      & =& \Pr(A=0|B=0) Pr(B=0)\\
                      &  & + \Pr(A=1|B=1) Pr(B=1)\\
                      & =& \frac{1}{2} [\Pr(A=0|B=0) + \Pr(A=1|B=1)]\\
                      &=& p(\gamma_1,\gamma_2)
\end{eqnarray*}}

Now,
{\scriptsize
\begin{eqnarray*}
\Pr (A=0|B=0) &
=&\frac{1}{3}[\Pr(C=\ket{\phi_0}|A'=\ket{1'},B'=\ket{0})\\
&  & + \Pr(C=\ket{\phi_0}| A'=\ket{0'}, B'=\ket{1})\\
&  & + \Pr(C=\ket{\phi_0}| A'=\ket{2'}, B'=\ket{1})\\
&  & + \Pr(C=\ket{\phi_0}| A'=\ket{0'}, B'=\ket{2})\\
&  & + \Pr(C=\ket{\phi_0}| A'=\ket{1'}, B'=\ket{2})]\\
& = & \frac{1}{6} (2\sin^2 \gamma_1 + 2\sin^2 \gamma_2 - \sin^2 \gamma_1 \sin^2 \gamma_2)
\end{eqnarray*}}
Where the event $C=\ket{\phi_0}$ implies Charlie measures $\ket{\phi_0}$. $A'$ and $B'$ represent the states sent by Alice and Bob respectively.

From Table $2$, it is clear that,
$\Pr(A=0|B=0) = \Pr(A=1|B=1)$
because of symmetry and thus,

 {\scriptsize
\begin{eqnarray*}
p(\gamma_1,\gamma_2)&=&\Pr(A=B) \\
&=& \frac{1}{6} (2\sin^2 \gamma_1 + 2\sin^2 \gamma_2 - \sin^2 \gamma_1 \sin^2 \gamma_2).
\end{eqnarray*}}
Figure $1$ shows the change in the database privacy with the change of $\gamma_1$ and $\gamma_2$.
\begin{figure}
\centering
\includegraphics[width=2.5in, height=3in]{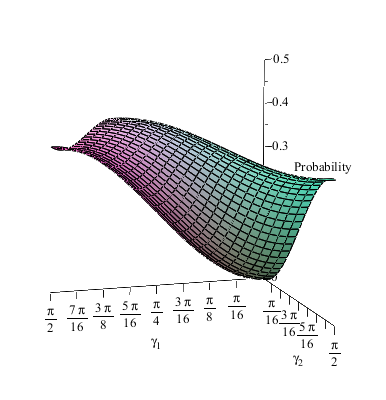}
\caption{Change in the database privacy with $\gamma_1$ and $\gamma_2$.}
\end{figure}
In case of MDI QPQ with qubits~\cite{zhao}, the success probability of Alice
to obtain a conclusive raw key is $p'(\theta)=\frac{\sin^2
\theta}{2}$, where the parameter $\theta \in (0,
\frac{\pi}{2})$ is chosen by Bob.

Note that in case of qutrit the success probability of Alice
contains two parameters $\gamma_1$, $\gamma_2$. However, in case of
qubit it contains only one parameter $\theta$. So, we cannot compare
our result with~\cite{zhao} in general. To show which one gives
better database privacy we have to keep at least one of the
parameters for qutrit fixed and compare another with $\theta$.

Let us define, $R_1:=\{\gamma_1,\gamma_2\in(0,
\frac{\pi}{2}):-\sin^2\gamma_1+2\sin^2\gamma_2-\sin^2\gamma_1\sin^2\gamma_2<0\}$
and $R_2:=\{\gamma_1,\gamma_2\in(0,
\frac{\pi}{2}):2\sin^2\gamma_1-\sin^2\gamma_2-\sin^2\gamma_1\sin^2\gamma_2<0\}$.
We can see that, in the region $R_1$,
$p(\gamma_1,\gamma_2)<p'(\gamma_1)$ and in the region $R_2$,
$p(\gamma_1,\gamma_2)<p'(\gamma_2)$. That means in these two regions Alice's success probability reduces if we consider qutrit. As the database is encrypted by the key possessed by Bob, thus if we can reduce the success probability of Alice to guess a key bit of Bob, we can increase the security of the database.

The MDI QPQ protocol using qutrits with parameters $(\gamma_1,
\gamma_2)\in R_1$ gives better database security than the MDI QPQ
protocol using qubits with the parameter $\theta=\gamma_1$.
Similarly the MDI QPQ protocol with qutrit and with $(\gamma_1,
\gamma_2)\in R_2$ gives better database security than the MDI QPQ
protocol with qubits and with $\theta=\gamma_2$. Figure $2$ shows the region for
$R_1$ and $R_2$ with red and blue shades respectively. In other
words, for the values of $(\gamma_1, \gamma_2)$ in the colored
region of the Figure 2, $3d$-MDI QPQ protocol gives better database
security than $2d$-MDI QPQ protocol performed with either
$\gamma_1$ or $\gamma_2$.

\begin{figure}
\centering
\includegraphics[width=2.5in, height=3in]{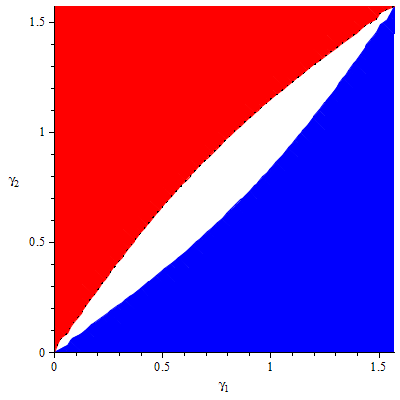}
\caption{Comparison of database privacy for 3D MDI QPQ and 2D MDI QPQ.}
\end{figure}

\subsection{User Privacy}
\label{attack}

User privacy is analyzed against Middle State Attack introduced in~\cite{Gao}. In the Middle State Attack it is assumed that Bob is dishonest. He tries to guess the position of the key bit that Alice has obtained with certainty.

Now, we will revisit~\cite{zhao} to understand how the Middle State Attack for MDI QPQ with qubits has been defended. At first, we will consider the case when both Bob and Alice are honest.

At the beginning of the protocol, Bob and Alice send one of the four polarized photons $\ket{0}, \ket{1}, \ket{0'}, \ket{1'}$ uniformly at random to the BSM possessed by Alice, where
\begin{eqnarray*}
\ket{0'} & = & \cos\theta\ket{0} + \sin\theta\ket{1}\\
\ket{1'} & = & \sin\theta\ket{0} - \cos\theta\ket{1}.
\end{eqnarray*}
 The rectilinear basis
$\{\ket{0}, \ket{1}\}$ is encoded to the key bit $0$ and the basis
$\{\ket{0'}, \ket{1'}\}$ is encoded to the key bit $1$. For each
measurement output $$\ket{\psi^-}
=\frac{1}{\sqrt{2}}(\ket{01}-\ket{10}),$$ Bob will declare a key bit
$0$ if he has sent $\ket{0}$ or $\ket{0'}$ and $1$ for $\ket{1}$ or
$\ket{1'}$. Probability of obtaining $\ket{\psi^-}$ for different
possibilities are shown in the Table $3$.
\begin{table}[htbp]
\centering
\begin{tabular}{|c|c|c|c|c|c|}
\hline
& \multicolumn{4}{c}{Bob} & \\
\cline{2-6}
\multirow{6}{*}{Alice} & & $\ket{0}$ & $\ket{1}$ & $\ket{0'}$ & $\ket{1'}$\\
\hline
 & $\ket{0}$ & 0 & $\frac{1}{2}$ & $\frac{1}{2}{\sin^2 \theta}$ & $\frac{1}{2}{\cos^2 \theta}$\\
\cline{2-6}
            & $\ket{1}$ & $\frac{1}{2}$ & 0 & $\frac{1}{2}{\cos^2\theta}$ & $\frac{1}{2}{\sin^2 \theta}$\\
\cline{2-6}
            & $\ket{0'}$ & $\frac{1}{2}{\sin^2 \theta}$ & $\frac{1}{2}{\cos^2\theta}$ & 0 & $\frac{1}{2}$\\
\cline{2-6}
            & $\ket{1'}$ & $\frac{1}{2}{\cos^2 \theta}$ & $\frac{1}{2}{\sin^2 \theta}$ & $\frac{1}{2}$ & 0 \\
\hline
\end{tabular}
\caption{Theoretical probabilities of obtaining $\ket{\psi^-}$ for different combination of honest states}
\end{table}

Based on the declaration of Bob and the qubit Alice has sent, Alice
can guess Bob's qubit with certainty. For example, suppose Bob has
sent $\ket{0}$ and declared $0$, then from Table $3$ it can be seen
that Alice guesses Bob's qubit only when she has sent $\ket{0'}$. In
this case she will conclude the key bit as $0$. The probability that
Alice can guess  a conclusive raw key bit is
$\frac{\sin^2\theta}{2}$.

For simplicity we consider that Alice and Bob choose the same $\theta$. Another variant of the
protocol is discussed in~\cite{zhao} where one of them uses $\theta$
and other one uses $(\theta+\frac{\pi}{2})$. The analysis is
same for both the variants.

In case of Middle State Attack a dishonest Bob sends a qubit
uniformly at random in the following form
\begin{eqnarray*}
\ket{0''} & = & \cos\frac{\theta}{2}\ket{0} + \sin\frac{\theta}{2}\ket{1}\\
\ket{1''} & = & \sin\frac{\theta}{2}\ket{0} - \cos\frac{\theta}{2}\ket{1}
\end{eqnarray*}
and declares $1$ if he has sent $\ket{0''}$ and $0$ if he has sent
$\ket{1''}$. Probabilities of obtaining $\ket{\psi^-}$ for different
possibilities in this case is shown in the Table $4$.
\begin{table}[htbp]
\centering
\begin{tabular}{|c|c|c|c|c|c|}
\hline
 & \multicolumn{4}{c}{Alice} & \\
\cline{2-6}
\multirow{4}{*}{Bob} & & $\ket{0}$ & $\ket{1}$ & $\ket{0'}$ & $\ket{1'}$\\
\hline
 & $\ket{0''}$ & $\frac{1}{2}{\sin^2\frac{\theta}{2}}$ & $\frac{1}{2}{\cos^2\frac{\theta}{2}}$ & $\frac{1}{2}{\sin^2\frac{\theta}{2}}$ & $\frac{1}{2}{\cos^2\frac{\theta}{2}}$\\
\cline{2-6}
            & $\ket{1''}$ & $\frac{1}{2}{\cos^2\frac{\theta}{2}}$ & $\frac{1}{2}{\sin^2\frac{\theta}{2}}$ & $\frac{1}{2}{\cos^2\frac{\theta}{2}}$ & $\frac{1}{2}{\sin^2\frac{\theta}{2}}$\\
\hline
\end{tabular}
\caption{Theoretical probabilities of obtaining $\ket{\psi^-}$ for different combination of middle states}
\end{table}

Let us consider the instance when Bob has sent $\ket{0''}$ and
declares $1$. In this case Alice will conclude her key bit as $0$ if she has
sent $\ket{1'}$ and $1$ if she has sent $\ket{1}$.

Now, the probability to guess a key bit as $0$ by Alice is
{\scriptsize
\begin{eqnarray*}
p_0^{qubit} & = & \Pr(BSM=\ket{\psi^-}|A'=\ket{1'}, B'=\ket{0''})\\
            & = & \frac{1}{2}\cos^2\frac{\theta}{2}
\end{eqnarray*}}
and the probability to guess a key bit as $1$ by Alice is
{\scriptsize
\begin{eqnarray*}
p_1^{qubit} & = & \Pr(BSM=\ket{\psi^-}|A'=\ket{1}, B'=\ket{0''})\\
            & = & \frac{1}{2}\cos^2\frac{\theta}{2}
\end{eqnarray*}}

Let $E$ be the event that Alice concludes the raw key bit  with
certainty. Hence, the total probability to get a conclusive raw key bit by Alice is
{\scriptsize
\begin{eqnarray*}
p_{c,mid}^{qubit} & = & \Pr(E)\\
                   & = & \Pr(E|B'=\ket{0''})\Pr(B'=\ket{0''})\\
                   &  & +\Pr(E|B'=\ket{1''})\Pr(B'=\ket{1''})\\
                   & = &
                   \frac{1}{2}[\Pr(E|B'=\ket{0''})+\Pr(E|B'=\ket{1''})]\\
                   & = & \frac{1}{2} [\Pr(BSM=\ket{\psi^-}|A'=\ket{1'}, B'=\ket{0''})\\
                   &  & + \Pr(BSM=\ket{\psi^-}|A'=\ket{1}, B'=\ket{0''})\\
                   &  & + \Pr(BSM=\ket{\psi^-}|A'=\ket{0'}, B'=\ket{1''})\\
                   &  & + \Pr(BSM=\ket{\psi^-}|A'=\ket{0},
                   B'=\ket{1''})]\\
                   & = & \cos^2\frac{\theta}{2}
\end{eqnarray*}}
Figure $3$ shows the success probability of Alice for honest and dishonest Bob in case of MDI QPQ with qubits.

From the Figure 3, it is clear that for $\theta \in (0,\frac{\pi}{2})$, $\cos^2\frac{\theta}{2}> \frac{1}{2}\sin^2{\theta}$. Thus, mounting the Middle State Attack Bob can increase the success probability of Alice to get a conclusive key bit.

The motivation of Bob to increase the success probability of Alice is to track the positions where there is a higher probability to get a conclusive key bits. However, he can not decide if it is $0$ or $1$ as
the success probability of Alice to get the key bit as $0$ and as $1$ is same.
In this case he has to insert the key bits randomly. And this will introduce an error in the channel. When Alice and Bob perform error correction this error is identified. Alice will identify this error as quantum bit error rate (QBER). In this case, QBER is equal to $\frac{1}{2}$. Noticing this Alice will abort the Protocol.

\begin{figure}
\centering
\includegraphics[width=2.5in, height=2.5in]{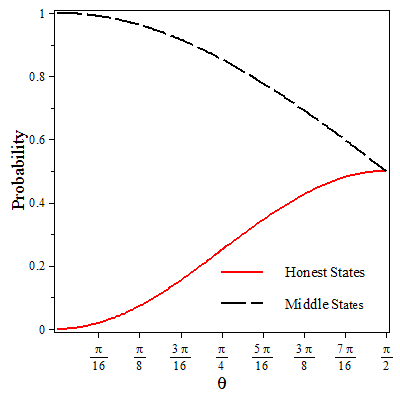}
\caption{Comparison in database privacy for honest Bob and dishonest Bob for qubits.}
\end{figure}

In case of qutrit, Bob will send one of the  following states
 uniformly at random to Charlie and announces $1$, $2$, $0$ respectively to Alice.

\begin{eqnarray*}
    \ket{0''} =& \cos \frac{\gamma_1}{2} \ket{0} + \sin \frac{\gamma_1}{2} \cos \frac{\gamma_2}{2} \ket{1} + \sin \frac{\gamma_1}{2} \sin \frac{\gamma_2}{2} \ket{2}, \\
    \ket{1''} =& -\sin \frac{\gamma_1}{2} \ket{0} + \cos \frac{\gamma_1}{2} \cos \frac{\gamma_2}{2} \ket{1} + \cos \frac{\gamma_1}{2} \sin \frac{\gamma_2}{2} \ket{2}, \\
    \ket{2''} =& -\sin \frac{\gamma_2}{2} \ket{1} + \cos \frac{\gamma_2}{2} \ket{2}
\end{eqnarray*}
Theoretical probabilities (after normalization) for obtaining the
Bell state $\ket{\phi_0}$ in this case are shown in the Table $5$.

\begin{table*}[htbp]
\centering
{\scriptsize
\begin{tabular}{|c|c|c|c|c|}
\hline
 & \multicolumn{3}{c}{Bob} & \\
\cline{2-5}
& & $\ket{0''}$ & $\ket{1''}$ & $\ket{2''}$\\
\hline
&&&&\\
\multirow{19}{*}{Alice} & $\ket{0}$ & $\frac{1}{2}{\cos^2{\frac{\gamma_1}{2}}}$ & $\frac{1}{2}\sin^2\frac{\gamma_1}{2}$ & 0\\
&&&&\\
\cline{2-5}
&&&&\\
& $\ket{1}$ & $\frac{1}{2}{\sin^2{\frac{\gamma_1}{2}}\cos^2\frac{\gamma_2}{2}}$ & $\frac{1}{2}{\cos^2\frac{\gamma_1}{2}\cos^2\frac{\gamma_2}{2}}$ & $\frac{1}{2}{\sin^2\frac{\gamma_2}{2}}$\\
&&&&\\
\cline{2-5}
&&&&\\
& $\ket{2}$ & $\frac{1}{2}{\sin^2\frac{\gamma_1}{2}\sin^2\frac{\gamma_2}{2}} $ & $\frac{1}{2}{\cos^2\frac{\gamma_1}{2}\sin^2\frac{\gamma_2}{2}} $ & $\frac{1}{2}{\cos^2\frac{\gamma_2}{2}} $\\
&&&&\\
\cline{2-5}
&&&&\\
& $\ket{0'}$ & $\frac{1}{2}(\cos\gamma_1\cos\frac{\gamma_1}{2} $ & $\frac{1}{2}(\cos\gamma_1\sin\frac{\gamma_1}{2} $ & $\frac{1}{2}{\sin^2\gamma_1\sin^2\frac{\gamma_2}{2}} $\\
&  & $ + \sin\gamma_1\sin\frac{\gamma_1}{2}\cos\frac{\gamma_2}{2})^2 $ & $ - \sin\gamma_1\cos\frac{\gamma_1}{2}\cos\frac{\gamma_2}{2})^2 $ &\\
&&&&\\
\cline{2-5}
&&&&\\
& $\ket{1'}$ & $\frac{1}{2}(\sin\gamma_1\cos\frac{\gamma_1}{2} $ & $\frac{1}{2}(\sin\gamma_1\sin\frac{\gamma_1}{2} $ & $\frac{1}{2}{\cos^2\gamma_1\sin^2\frac{\gamma_2}{2}} $\\
&  & $ - \cos\gamma_1\sin\frac{\gamma_1}{2}\cos\frac{\gamma_2}{2})^2 $ & $ + \cos\gamma_1\cos\frac{\gamma_1}{2}\cos\frac{\gamma_2}{2})^2 $ &\\
&&&&\\
\cline{2-5}
&&&&\\
& $\ket{2'}$ & $ \frac{1}{2}{\sin^2\frac{\gamma_1}{2}\sin^2\frac{\gamma_2}{2}} $ & $ \frac{1}{2}{\cos^2\frac{\gamma_1}{2}\sin^2\frac{\gamma_2}{2}} $ & $\frac{1}{2}{\cos^2\frac{\gamma_2}{2}} $\\
&&&&\\
\hline
\end{tabular}}
\caption{Theoretical probabilities for obtaining Bell state $\ket{\phi_0}$ for dishonest Bob and honest Alice}
\end{table*}

Let us now consider the instance when Bob sends $\ket{0''}$ to
Charlie and declare the trit $1$. As Alice is honest, Alice will
conclude her raw key bit as $0$ if she has sent $\ket{0'}$
or $\ket{2'}$ and $1$ if she has sent $\ket{0}$ or $\ket{2}$.
Similar thing happens for other cases also.

Let $E$ be the event that Alice
concludes the raw key bit  with certainty. Then the total probability of obtaining a
conclusive raw key bit by Alice is
{\scriptsize
\begin{eqnarray*}
p_{c}^{mid} & = & \Pr(E)\\
            & = & \Pr(E|B'=\ket{0''})Pr(B'=\ket{0''})\\
            &  & + \Pr(E|B'=\ket{1''})Pr(B'=\ket{1''})\\
            &  & + \Pr(E|B'=\ket{2''})Pr(B'=\ket{2''})\\
            & = &\frac{1}{3}[Pr(E|B'=\ket{0''}) + Pr(E|B'=\ket{1''}) +
            Pr(E|B'=\ket{2''})]\\
            & = &\frac{1}{3}[Pr(C=\ket{\phi_0}|A'=\ket{0}, B'=\ket{0''})\\
            &  & +\Pr(C=\ket{\phi_0}|A'=\ket{2}, B'=\ket{0''})\\
            &  & +\Pr(C=\ket{\phi_0}|A'=\ket{0'}, B'=\ket{0''})\\
            &  & +\Pr(C=\ket{\phi_0}|A'=\ket{2'}, B'=\ket{0''})\\
            &  & +\Pr(C=\ket{\phi_0}|A'=\ket{1}, B'=\ket{1''})\\
            &  & +\Pr(C=\ket{\phi_0}|A'=\ket{0'}, B'=\ket{1''})\\
            &  & +\Pr(C=\ket{\phi_0}|A'=\ket{1'}, B'=\ket{1''})\\
            &  & +\Pr(C=\ket{\phi_0}|A'=\ket{1}, B'=\ket{2''})\\
            &  & +\Pr(C=\ket{\phi_0}|A'=\ket{2}, B'=\ket{2''})\\
            &  & +\Pr(C=\ket{\phi_0}|A'=\ket{1'},B'=\ket{2''})]\\
            & = &\frac{1}{6}[2+2\sin^2\frac{\gamma_1}{2}\sin^2\frac{\gamma_2}{2}+2\cos^2\frac{\gamma_1}{2}\cos^2\frac{\gamma_2}{2}\\
            &  & +\cos^2\gamma_1\sin^2\frac{\gamma_2}{2}+(\cos\gamma_1\cos\frac{\gamma_1}{2}+\sin\gamma_1\sin\frac{\gamma_1}{2}\cos\frac{\gamma_2}{2})^2]
\end{eqnarray*}}

Figure $4$ shows the comparison in the probability of obtaining a
raw key bit by Alice when Bob performs the protocol honestly with the probability of obtaining a
raw key bit by Alice when Bob performs the protocol dishonestly. Black surface shows the variation in
probability with respect to $\gamma_1$ and $\gamma_2$ for honest Bob
and the yellow surface shows the same when Bob sends middle
states. It is clear from the figure that, the probability to get a
key bit by Alice is very high in case of middle state than the honest
performance of the protocol.
\begin{figure}[ht]
\centering
\includegraphics[width=2.5in, height=3in]{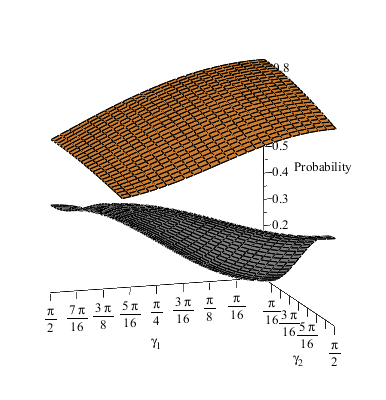}
\caption{Comparison in database privacy for honest Bob and dishonest Bob for qutrits.}
\end{figure}

\begin{comment}
Probability to get a conclusive raw
key by Alice in this case is given by

{\scriptsize
\begin{eqnarray*}
p_{inc}^{mid} & = & \Pr(E)\\
            & = & \Pr(E|B'=\ket{0''})Pr(B'=\ket{0''}) + \Pr(E|B'=\ket{1''})Pr(B'=\ket{1''})\\
            &  & + \Pr(E|B'=\ket{2''})Pr(B'=\ket{2''})\\
            & = &\frac{1}{3}[\Pr(E|B'=\ket{0''}) + \Pr(E|B'=\ket{1''}) +
            \Pr(E|B'=\ket{2''})]\\
            & = &\frac{1}{3}[\Pr(C=\ket{\phi_0}|A'=\ket{1}, B'=\ket{0''})\\
            &  & +\Pr(C=\ket{\phi_0}|A'=\ket{2}, B'=\ket{0''})\\
            &  & +\Pr(C=\ket{\phi_0}|A'=\ket{1'},B'=\ket{0''})\\
            &  & +Pr(C=\ket{\phi_0}|A'=\ket{0}, B'=\ket{1''})\\
            &  & +\Pr(C=\ket{\phi_0}|A'=\ket{2}, B'=\ket{1''})\\
            &  & +\Pr(C=\ket{\phi_0}|A'=\ket{0'}, B'=\ket{1''})\\
            &  & +\Pr(C=\ket{\phi_0}|A'=\ket{2'}, B'=\ket{1''})\\
            &  & +\Pr(C=\ket{\phi_0}|A'=\ket{1}, B'=\ket{2''})\\
            &  & +\Pr(C=\ket{\phi_0}|A'=\ket{0'}, B'=\ket{2''})\\
            &  & +\Pr(C=\ket{\phi_0}|A'=\ket{1'}, B'=\ket{2''})]\\
            & = &\frac{1}{6}[2\sin^2\frac{\gamma_1}{2}+4\sin^2\frac{\gamma_2}{2}-2\sin^2\frac{\gamma_1}{2}\sin^2\frac{\gamma_2}{2}\\
            &  & +(\sin\gamma_1\cos\frac{\gamma_1}{2}-\cos\gamma_1\sin\frac{\gamma_1}{2}\cos\frac{\gamma_2}{2})^2\\
            &  & +(\cos\gamma_1\sin\frac{\gamma_1}{2}-\sin\gamma_1\cos\frac{\gamma_1}{2}\cos\frac{\gamma_2}{2})^2]
\end{eqnarray*}}
\end{comment}

Now, let us consider the example when Bob sends $\ket{0''}$ and declares
$1$. In this case Alice will conclude the key bit as $0$ if she has sent $\ket{0'}$
or $\ket{2'}$ and $1$ if she has sent $\ket{0}$ or $\ket{2}$. The probability to get conclusive raw key bit as $0$ is
{\scriptsize
\begin{eqnarray*}
p_{0} & = & \Pr(C=\ket{\phi_0}|A'=\ket{0'},B'=\ket{0''})\\
      &  & +\Pr(C=\ket{\phi_0}|A'=\ket{2'},B'=\ket{0''})\\
      & = &\frac{1}{2}((\cos\gamma_1\cos\frac{\gamma_1}{2}+\sin\gamma_1\sin\frac{\gamma_1}{2}\cos\frac{\gamma_2}{2})^2+\sin^2\frac{\gamma_1}{2}\sin^2\frac{\gamma_2}{2})
\end{eqnarray*}}
Similarly, the probability to get conclusive raw key bit as $1$ is
{\scriptsize
\begin{eqnarray*}
p_{1} & = & \Pr(C=\ket{\phi_0}|A'=\ket{0},B'=\ket{0''})\\
      &  & + \Pr(C=\ket{\phi_0}|A'=\ket{2},B'=\ket{0''})\\
      & = &\frac{1}{2}(\cos^2\frac{\gamma_1}{2}+\sin^2\frac{\gamma_1}{2}\sin^2\frac{\gamma_2}{2}).
\end{eqnarray*}}

%\begin{widetext}
Now,
{\scriptsize
\begin{eqnarray*}
 &\cos\frac{\gamma_2}{2}<1,\\
 &\Rightarrow  \sin\gamma_1\sin\frac{\gamma_1}{2}\cos\frac{\gamma_2}{2}<\sin\gamma_1\sin\frac{\gamma_1}{2}\\
& \Rightarrow  \cos\gamma_1\cos\frac{\gamma_1}{2}+\sin\gamma_1\sin\frac{\gamma_1}{2}\cos\frac{\gamma_2}{2}<\cos\gamma_1\cos\frac{\gamma_1}{2}+\sin\gamma_1\sin\frac{\gamma_1}{2}\\
 &\Rightarrow  (\cos\gamma_1\cos\frac{\gamma_1}{2}+\sin\gamma_1\sin\frac{\gamma_1}{2}\cos\frac{\gamma_2}{2})^2
 <\cos^2\frac{\gamma_1}{2}\\
 &\Rightarrow \frac{1}{2}[(\cos\gamma_1\cos\frac{\gamma_1}{2}+\sin\gamma_1\sin\frac{\gamma_1}{2}\cos\frac{\gamma_2}{2})^2+\sin^2\frac{\gamma_1}{2}\sin^2\frac{\gamma_2}{2}]
 < \frac{1}{2}({\cos^2\frac{\gamma_1}{2}+\sin^2\frac{\gamma_1}{2}\sin^2\frac{\gamma_2}{2}})\\
& \Rightarrow  p_0<p_1,
\end{eqnarray*}}
for all $\gamma_1, \gamma_2 \in (0, \frac{\pi}{2})$.

%\end{widetext}

As $p_0<p_1$, with a very high probability Alice will conclude the
key bit as $1$. Bob also encodes this event with the key bit $1$.
And hence, the key bit of Alice and the key bit of Bob matches with
high probability.

Now with this Middle State Attack Bob can guess the position $j$ of Alice's key bit with a very high probability. When Alice sends $s=j-i$, Bob will immediately come to know the position $i$ of the the database element as $i=j-s$.

For the other two declarations, i.e., for $2$ and $0$, the same thing happens.

The attack can be defended if we set threshold for error probability at error correcting phase strictly less than $p_0$. This is because, with probability $p_0$, Alice sets the key bit as $0$. However, in that case, Bob will set the key bit as $1$. This error is identified in error correcting phase. If the threshold value for error probability is higher than $p_0$, Bob will pass the test and can mount the attack successfully. On the other hand, setting the threshold value strictly below $p_0$, one can resist Bob to mount the attack.

Another way to defend this attack is to set $p_0=p_1$ i.e., when
$\gamma_2=0$. In that case, the second basis states becomes

\begin{eqnarray*}
    \ket{0'} =& \cos \gamma_1 \ket{0} + \sin \gamma_1 \ket{1}, \\
    \ket{1'} =& -\sin \gamma_1 \ket{0} + \cos \gamma_1 \ket{1}, \\
    \ket{2'} =& \ket{2}
\end{eqnarray*}

This case is similar to MDI QPQ with qubits~\cite{zhao}.
\section{A Spacial Case}
In this section we will discuss MDI QPQ protocol for Fourier
basis. The Fourier basis states are as follows.

\begin{eqnarray*}
     \ket{0'}=\frac{1}{\sqrt{3}}(\ket{0}+\ket{1}+\ket{2}),\\
     \ket{1'}=\frac{1}{\sqrt{3}}(\ket{0}+\omega^2\ket{1}+\omega\ket{2}),\\
     \ket{2'}=\frac{1}{\sqrt{3}}(\ket{0}+\omega\ket{1}+\omega^2\ket{2}).
\end{eqnarray*}
where $\omega = e^{\frac{2\pi i}{3}}$.

 The state $\ket{0'}$
belongs to a specific ensemble with $\gamma_2 = \frac{\pi}{4}$ and
 $\cos \gamma_1 = \frac{1}{\sqrt{3}}$.
The set $\{\ket{0'}, \ket{1'},\ket{2'}\}$ form an orthonormal basis and is related
to the computational basis by Discrete Fourier Transform.

Now the $3d$-MDI QPQ protocol will be performed with this
new set of basis. For the earlier case, Alice and Bob have to consider the cases where
 the measurement output was $\ket{\phi_0}$, whereas for this case they can choose any of the nine Bell states.
Theoretical probabilities for obtaining $\ket{\phi_0}$ for different
combinations are shown in the Table $6$.
\begin{table}[htbp]
\centering
\begin{tabular}{|c|c|c|c|c|c|c|c|}
\hline
 & \multicolumn{6}{c}{Bob} & \\
\cline{2-8}
& & $\ket{0}$ & $\ket{1}$ & $\ket{2}$ & $\ket{0'}$ & $\ket{1'}$ & $\ket{2'}$\\
\hline
\multirow{6}{*}{Alice} & $\ket{0}$ & $\frac{1}{3}$ & 0 & 0 & $\frac{1}{9}$ & $\frac{1}{9}$ & $\frac{1}{9}$\\
\cline{2-8}
            & $\ket{1}$ & 0 & $\frac{1}{3}$ & 0 & $\frac{1}{9}$ & $\frac{1}{9}$ & $\frac{1}{9}$\\
\cline{2-8}
            & $\ket{2}$ & 0 & 0 & $\frac{1}{3}$ & $\frac{1}{9}$ & $\frac{1}{9}$ & $\frac{1}{9}$\\
\cline{2-8}
            & $\ket{0'}$ & $\frac{1}{9}$ & $\frac{1}{9}$ & $\frac{1}{9}$ & $\frac{1}{3}$ & 0 & 0 \\
\cline{2-8}
            & $\ket{1'}$ & $\frac{1}{9}$ & $\frac{1}{9}$ &  $\frac{1}{9}$ & 0 & $\frac{1}{3}$ & 0 \\
\cline{2-8}
            & $\ket{2'}$ & $\frac{1}{9}$ & $\frac{1}{9}$ & $\frac{1}{9}$ & 0 & 0 & $\frac{1}{3}$\\
\hline
\end{tabular}
\caption{Theoretical probabilities for obtaining Bell state $\ket{\phi_0}$ for different combination of states}
\end{table}

\begin{table}[htbp]
\centering
{\scriptsize
\begin{tabular}{|c|c|c|c|c|c|c|c|}
\hline
 & \multicolumn{6}{c}{Bob} & \\
\cline{2-8}
& & $\ket{0}$ & $\ket{1}$ & $\ket{2}$ & $\ket{0'}$ & $\ket{1'}$ & $\ket{2'}$\\
\hline
\multirow{6}{*}{Alice} & $\ket{0}$ & $\frac{1}{2}$ & 0 & 0 & $\frac{1}{6}$ & $\frac{1}{6}$ & $\frac{1}{6}$\\
\cline{2-8}
            & $\ket{1}$ & 0 & $\frac{1}{2}$ & 0 & $\frac{1}{6}$ & $\frac{1}{6}$ & $\frac{1}{6}$\\
\cline{2-8}
            & $\ket{2}$ & 0 & 0 & $\frac{1}{2}$ & $\frac{1}{6}$ & $\frac{1}{6}$ & $\frac{1}{6}$\\
\cline{2-8}
            & $\ket{0'}$ & $\frac{1}{6}$ & $\frac{1}{6}$ & $\frac{1}{6}$ & $\frac{1}{2}$ & 0 & 0 \\
\cline{2-8}
            & $\ket{1'}$ & $\frac{1}{6}$ & $\frac{1}{6}$ &  $\frac{1}{6}$ & 0 & $\frac{1}{2}$ & 0 \\
\cline{2-8}
            & $\ket{2'}$ & $\frac{1}{6}$ & $\frac{1}{6}$ & $\frac{1}{6}$ & 0 & 0 & $\frac{1}{2}$\\
\hline
\end{tabular}}
\caption{Theoretical probabilities for obtaining Bell state $\ket{\phi_0}$ for different combination of states after normalization}
\end{table}

The protocol remains same as earlier. To avoid repetition we do not
write the whole protocol once again. From Table $7$, we can
calculate the success probability of Alice to obtain a conclusive
key bit.

{\scriptsize
\begin{eqnarray*}
p & =& \Pr (A=B)\\
  & =& \Pr(A=0, B=0) + \Pr(A=1, B=1)\\
  & =& \Pr(A=0|B=0) \Pr(B=0)\\
  &  & + \Pr(A=1|B=1) \Pr(B=1)\\
  & =& \frac{1}{2} [\Pr(A=0|B=0) + \Pr(A=1|B=1)]
\end{eqnarray*}
}
Now,
{\scriptsize
\begin{eqnarray*}
\Pr (A=0|B=0) & = & \frac{1}{3}[\Pr(C=\ket{\phi_0}| A'=\ket{1'},B'=\ket{0})\\
             &  & + \Pr(C=\ket{\phi_0}| A'=\ket{2'},B'=\ket{0})\\
             &  & + \Pr(C=\ket{\phi_0}| A'=\ket{0'}, B'=\ket{1})\\
             &  & + \Pr(C=\ket{\phi_0}| A'=\ket{2'}, B'=\ket{1})\\
             &  & + \Pr(C=\ket{\phi_0}| A'=\ket{0'}, B'=\ket{2})\\
             &  & + \Pr(C=\ket{\phi_0}| A'=\ket{1'}, B'=\ket{2})]\\
             & = & \frac{1}{3}(6\times \frac{1}{6})\\
             & = & \frac{1}{3}
\end{eqnarray*}
}
As  $\Pr(A=0|B=0) = \Pr(A=1|B=1)$, we can write,
{\scriptsize
\begin{eqnarray*}
p & =& \Pr(A=B)\\
  & =& \frac{1}{3}
\end{eqnarray*}
}

In case of qubits the success probability of Alice to
get a conclusive key bit in Fourier basis is $\frac{1}{4}$~\cite{jakobi}. Contrary to this, here it becomes $\frac{1}{3}$. Thus, in this case we will not obtain better database security. And hence, we do not bother about the Middle State Attack.

\section{Conclusion}
It has been identified that in Quantum Key Distribution higher dimension provides better security. In~\cite{jo} the enhancement in the key rate of MDI QKD has been achieved exploiting qutrit. Motivated by this we try to understand whether this is the case for general cryptographic primitives.

In this direction, we explore MDI QPQ with qutrit. Our analysis shows a counter intuitive result. We observe that though database security can be enhanced using qutrit, the client's privacy becomes vulnerable. We can not optimize both the database and client security in three dimension simultaneously.
Hence, we conclude that higher dimension is not always advantageous in case of cryptographic primitives. It varies from protocol to protocol.

In the present draft we explore MDI-QPQ only. What happens for other QPQ protocols in three dimension would be very interesting cryptanalysis. Analysis of security issues for MDI QPQ and other QPQ protocols for greater than $3$ dimension is our future research plan.\\

\ \\
{\bf{Acknowledgments:}} The authors like to thank the anonymous reviewer for excellent comments that substantially improved the editorial as well as technical presentation of this paper.\\
\ \\

%\tableofcontents  % optional

%\markboth{Authors' Names}
%{Instructions for Typing Manuscripts (Paper's Title)}

%\section{References}

\end{document}